\documentclass[twocolumn,showpacs,preprintnumbers,amsmath,amssymb,superscriptaddress]{revtex4}

\usepackage{graphicx}
\usepackage{subfig}
\usepackage{epsfig}

\newcommand{\be}{\begin{equation}}
\newcommand{\ee}{\end{equation}}
\newcommand{\bea}{\begin{eqnarray}}
\newcommand{\eea}{\end{eqnarray}}
\newcommand{\ba}{\begin{array}}
\newcommand{\ea}{\end{array}}

\begin{document}


\title{Directional motion of forced polymer chains with hydrodynamic interaction}

\author{D. Santos-Oliv\'an}

\affiliation{Departamento de F\'{\i}sica de la Materia Condensada,
Universidad de Zaragoza, E-50009 Zaragoza, Spain}

\affiliation{Instituto de Biocomputaci\'on y F\'{\i}sica de Sistemas
Complejos, Universidad de Zaragoza, E-50009 Zaragoza, Spain}

\author{A. Fiasconaro}
\email{afiascon@unizar.es}
\affiliation{Departamento de F\'{\i}sica de la Materia Condensada,
Universidad de Zaragoza, E-50009 Zaragoza, Spain}

\affiliation{Instituto de Ciencia de Materiales de Arag\'on,
C.S.I.C.-Universidad de Zaragoza, E-50009 Zaragoza, Spain.}

\author{F. Falo}

\affiliation{Departamento de F\'{\i}sica de la Materia Condensada,
Universidad de Zaragoza, E-50009 Zaragoza, Spain}

\affiliation{Instituto de Biocomputaci\'on y F\'{\i}sica de Sistemas
Complejos, Universidad de Zaragoza, E-50009 Zaragoza, Spain}

\date{\today}

\begin{abstract}
We study the propulsion of a one-dimensional (1D) polymer chain under sinusoidal external forces in the overdamped (low Reynolds number)
regime. We show that, when hydrodynamical interactions are included, the polymer presents directional motion which depends on the phase differences of the external force applied along the chain. Moreover, the velocity shows a maximum as a function of the frequency. We discuss the relevance of all these results in light of recent nanotechnology experiments.
\end{abstract}


\maketitle

\section{Introduction}

Propulsion at the low Reynolds number regime is an interesting subject of investigation which induces us to review our intuition related to motion. The absence of inertial effects in this regime, along with the reduced number of degrees of freedom in small swimmers, makes the possibility of movement not obvious at all. At this regime, the hydrodynamical interactions takes the essential importance being able to give the non reciprocity necessary to the net movement of a structured object. After the seminal work of Purcell \cite{purcell}, and the one of Shapere and Wilczek \cite{shapere1987}, an increasing interest has recently arisen and a good number of papers have appeared in the last decade to describe self propelled swimmers with, at least, two degrees of freedom, the minimal condition to guarantee the movement at low Reynolds numbers\cite{purcell}. The model proposed by Purcell, consisting of three joined rods, has been studied recently by Tam \textit{et al}. \cite{tam2007}.
Other models consider the non reciprocal movement of two spheres joined together by an extensible bond that are able to change their volumes \cite{avron2005}. Among the simplest models, one consisting of three spheres linearly joined together by means of extensible bonds \cite{earl2007,3_esferas_1,3_esferas_2} has been also proposed. Those one dimensional systems are actually the simplest possible devices with two degrees of freedom, where the motion is controlled by the phase difference between the relative motion of the spheres. An extension of the three (and more) particle model to two dimensions was studied by the authors of Ref. \cite{earl2007}.

All these models deal mainly with the problem of self-propulsion, that is, the motion is caused by periodic deformation of their structure (see also Ref.~\cite{LaugaRPP} for a comprehensive review of these models), while in this paper we address a more general problem: the propulsion of a structured object (a polymer chain) submitted to non-homogenous time dependent driving forces. Recently, experiments at nanometric scale have been performed by Dreyfus \textit{et al}. \cite{dreyfus2005,dreyfus-nat2005}. In such works magnetic spheres joined by DNA links are driven by external fields. The velocity as a function of the frequency for this system shows a clear maximum. Moreover, a new experiment using also oscillating magnetic fields driving a macroscopic extended body has been carried out by Garstecki \emph{et al.} \cite{garstecki2009}, finding again a nonmonotonic dependence of the velocity \emph{versus} the frequency of the applied field. Polymeric models that may qualitatively explain the experimental results have been implemented in Ref.~\cite{earl2007,gauger2006}

The system here studied presents rich and interesting properties. In fact, under this model, it is possible to obtain the above nonmonotonic behavior of the net velocity as a function of the frequency of the driving field. Moreover, a strong dependence of the velocity with the phase difference of the force is present, showing that this phase represents a control parameter able to drive the motion in one direction or in the other, and also to modulate the maximum possible value of the velocity.

\section{Model}

\begin{figure}[b]
 \begin{center}
\includegraphics[width=8.7cm]{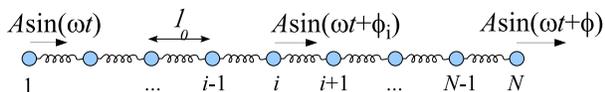}
\end{center}
\caption{(Color online) Scheme of the chain compound by $N$ monomers. The forces act on every sphere and the last one is shifted in phase with respect to the first one by a phase $\phi$. The $i$-th sphere presents a phase shift $\phi_i = i\frac{\phi}{N-1}$.}
\label{chain}

\end{figure}

\emph{Model equations.} The model consists of $N$ rigid coupled spheres joined to each other by a spring which can move in 1D along its $x$ axis (see Fig~\ref{chain}).

The system is supposed to be a few micrometers in size. By considering this scale and the typical velocities of microorganisms (or microscale or nanoscale devices), we obtain a Reynolds number around $R \approx 10^{-5}$. For this reason, we can use overdamped dynamics in our equations as a good approximation. At this scale we have to take into account the mutual interaction between particles due to the fluid motion (i.e., the hydrodynamic interaction). For small particles moving in a fluid, the Oseen tensor approximation \cite{Doi}, that relates the forces acting in each sphere with the respective velocities, can be used.
It is worth noting that in overdamped dynamics inertial effects are neglected and the system is not subject to acceleration. Consequently, in this approximation, the velocities in each instant depends on the instantaneous forces (through the Oseen tensor), and not on the past history of the system.

Let $\tilde{v}_i$ and $\tilde F_i$ be the velocity and force on the particle $i$, respectively. In one dimension, the Oseen tensor reduces to a square matrix with components $\tilde{H}_{ij}$ which relates $\tilde v_i$ to $\tilde F_i$,

\begin{equation}
\tilde v_i = \sum_{j=0}^{N-1} \tilde{H}_{ij} \tilde F_j.
  \label{relacion_fuerza_velocidad_3D}
\end{equation}

The components of the Oseen matrix  $\mathbf{\tilde{H}}$ are given by
\begin{equation}
  \tilde H_{ij} = \left\{ \begin{array}{c c} \frac{1}{6 \pi\eta a}, & \qquad i=j, \\
     \frac{1}{4 \pi \eta} \frac{1}{|x_i-x_j|} ,  & \qquad  i\neq j, \end{array}\right.\\
 \label{tensor_ossen}
\end{equation}
where $a$ and $x_i$ stands for the radius and position of particle $i$, respectively, and $\eta$ is the fluid viscosity.

In many of the previous works, and in particular in the three linked-spheres model \cite{3_esferas_1, 3_esferas_2}, the authors used internal forces to model the internal deformation of the system. Here, we drive the chain by an external force sinusoidal in time:

\begin{equation}
 K_i = \tilde A_i \sin (\omega t + \phi_i),
 \label{Fext}
\end{equation}
where the amplitudes $\tilde A_i$ are constant and equal for all the particles $\tilde A_i=\tilde A$, with $i=0,..., (N-1)$ and  $\phi$ is the total phase difference between the first and the last particle. The intermediate phases between the spheres are equally distributed along the chain, so that $\phi_i = i\frac{\phi}{N-1}$, with $i=0,..., (N-1)$.

Therefore, our system has three free parameters: the amplitude $\tilde A$, the frequency $\omega$ of the driving forces and the total phase difference $\phi$ along the chain.

The total force acting on the $i$-th particle, if we exclude the hydrodynamical contributions, is then:
\begin{equation}
  \tilde F_i = \tilde A_i \mathrm{sin}(\omega t + \phi_i) + k \Delta l_{i+} -  k \Delta l_{i-} \;\;\;i\in [0,...,N-1]\\
\end{equation}
where we define $\Delta l_{i+}$ as $x_i-x_{i-1} -l_0$ and $\Delta l_{i-}$ as $x_{i+1}-x_{i} -l_0$, $l_0$ being the natural length of the link. Both terms represent, respectively, the elongation of the chain along the bonds of the sphere $i$ with its neighbor. The parameter $k$ is the elastic constant of the spring which describes the harmonic interaction between spheres. The hydrodynamical approximation under which the above equation maintains its validity is that at any time the distance between the spheres $d$ is much higher than their radius $a \ll d$. Thus, to fulfill this condition we used $l_0=40a \gg a$.

It is important to note that the external contribution (\ref{Fext}) is in general different from zero [$\sum_i  K_i(t) \neq 0 $], while the long time average of the external force is null  [$\langle \sum_i K_i(t) \rangle _t = 0 $].

The choice of a linear dependence instead of other more complex functions appears to be not crucial for obtaining the net movement given by our model. In fact we calculated the velocity with other distributions of phases and they gave the same qualitative dependence, and also the net flux for zero force as present in other studies in the literature\cite{3_esferas_1}. We choose this linear distribution of the frequency essentially for two reasons: i) because it is simple and ii) because it is very easy to generalize it for an arbitrary number of spheres and to more complex (and realistic) systems of particles.

For calculation purposes, we use an adimensional set of equations, where the length unit is the radius of one sphere ($a$),  the time unit is  $6 \pi \eta a /k$, and the forces are measured in units of $ka$. The equations of motion obtained are then:

\begin{equation}
  \dot{x}_i(t) =  v_i = \sum_{j=0}^{N-1} H_{ij}(t) F_j(t),
  \label{eq-motion}
\end{equation}
where the new mobility matrix is given by:
\begin{equation}
H_{ij}(t) = \left\{ \begin{array}{c c}
		1, & \qquad i=j, \\
	      \frac{3}{2} \frac{1}{|x_i(t)-x_j(t)|},   & \qquad  i\neq j.
	       \end{array}\right.\\
\label{tensor_ossen_ad}
 \end{equation}

The system is not solvable analytically for an arbitrary number $N$ of particles. We solve numerically the equations using a Runge-Kutta algorithm of the fourth order, by using a variable increment $dt$, adjusted for the different frequencies of the driving force.

The main measure under study here is the asymptotic velocity of the polymer center of mass, and more precisely the mean velocity over a long time $T_M$, which is much bigger than the period of the applied force $T=2\pi / \omega$.
\begin{equation}
  v = \lim_{T_M \rightarrow \infty} \frac{1}{NT_M} \int_{t_0}^{t_0 + T_M} \sum_{j=0}^{N-1}{v_i},
\end{equation}
where $t_0$ is here a proper transient time in the dynamics.
This velocity depends on the total phase difference $\phi$ and on the frequency $\omega$ of the applied force. Although we have not been able to demonstrate the unicity of the velocity, the simulations (see below) indicate that this is the case independently on the initial conditions.

\vskip 0.2cm
\emph{Model symmetries.} The system under study presents interesting symmetries useful to demonstrate some of the features found in the numerical simulations.
We first note that it is simple to see that the equations of motion are periodic in the total phase difference with period $\phi_P = (N-1) 2\pi$. In fact, every sphere is submitted to a force $A \sin(\omega t + \phi_i)$, and so, if we add a phase  $\phi_P$ to the total phase difference, (i.e., $\phi \rightarrow \phi + \phi_P$), we obtain
\bea
 &  A \sin(\omega t + \phi_i) \rightarrow A \sin(\omega t + i \frac{\phi}{N-1} + 2\pi i) \nonumber \\ \nonumber
 & =A \sin(\omega t + \phi_i + 2\pi i)= A \sin(\omega t + \phi_i) ,
\eea
valid for any $\phi_i$. The first property is then:
\be
 v(\phi + \phi_P) = v(\phi).
 \label{p1}
\ee

On the other hand, we can also see that with the transformation $\phi_i \rightarrow -\phi_i$ the velocity changes sign. In fact, by using the time transformation the  $t \rightarrow -t \pm T/2$, we have
 \bea
 &  A \sin(\omega t + \phi_i) \rightarrow A \sin(-\omega t \pm \omega T/2+ \phi_i) = \nonumber \\ \nonumber
 & = -A \sin(\omega t - \phi_i \pm \pi) = A \sin(\omega t - \phi_i). \\ \nonumber
 \eea
This transformation maps the equation of motion for particles with phase $\phi$ to that one with phase $-\phi$ and opposite velocity. In fact, Eq.~(\ref{eq-motion}) can be formally written as:
 \be
 \frac{dx_i}{dt} = f[x_i, \sin(\omega t + \phi_i)].
 \ee
Using the above transformation in time ($t \rightarrow -t \pm T/2$), and the demonstrated \emph{sine} transformation, the transformed equation reads:
 \be
 -\frac{dx_i}{dt} = f[x_i, \sin(\omega t - \phi_i)].
 \ee

This equation is identical to the original one except for a global sign and for a change of the sign in all the phases. In other words, the second property is:
 \be
 -v(\phi) = v(-\phi).
 \label{p2}
 \ee

Note that this property allows us to choose the direction of the velocity by changing the relative phase between particles of the external driving forces.

By using the property (\ref{p2}) it is immediate to state that $v(0)=0$. Remembering also the property (\ref{p1}), it is also immediate that $v(\phi_P)=0$. Moreover, by using the property (\ref{p1}) with $\phi = -\phi_P/2$, we have:
 \be
   v(\phi_P/2) = v(-\phi_P/2).
 \ee
But using again the property (\ref{p2}) it is also true that
 \be
   v(\phi_P/2) = -v(-\phi_P/2).
 \ee
So it comes out that $v(\phi_P/2) = 0$.

In synthesis, we can summarize these last outcomes as
 \be
  v(0) = v(\phi_P/2) = v(\phi_P) = 0.
  \label{zero}
 \ee

\section{\label{results_2} Simulations.}
 \begin{figure}[tb]

  \subfloat[$N=2$]{
        \includegraphics[angle=-90,width=0.5\linewidth]{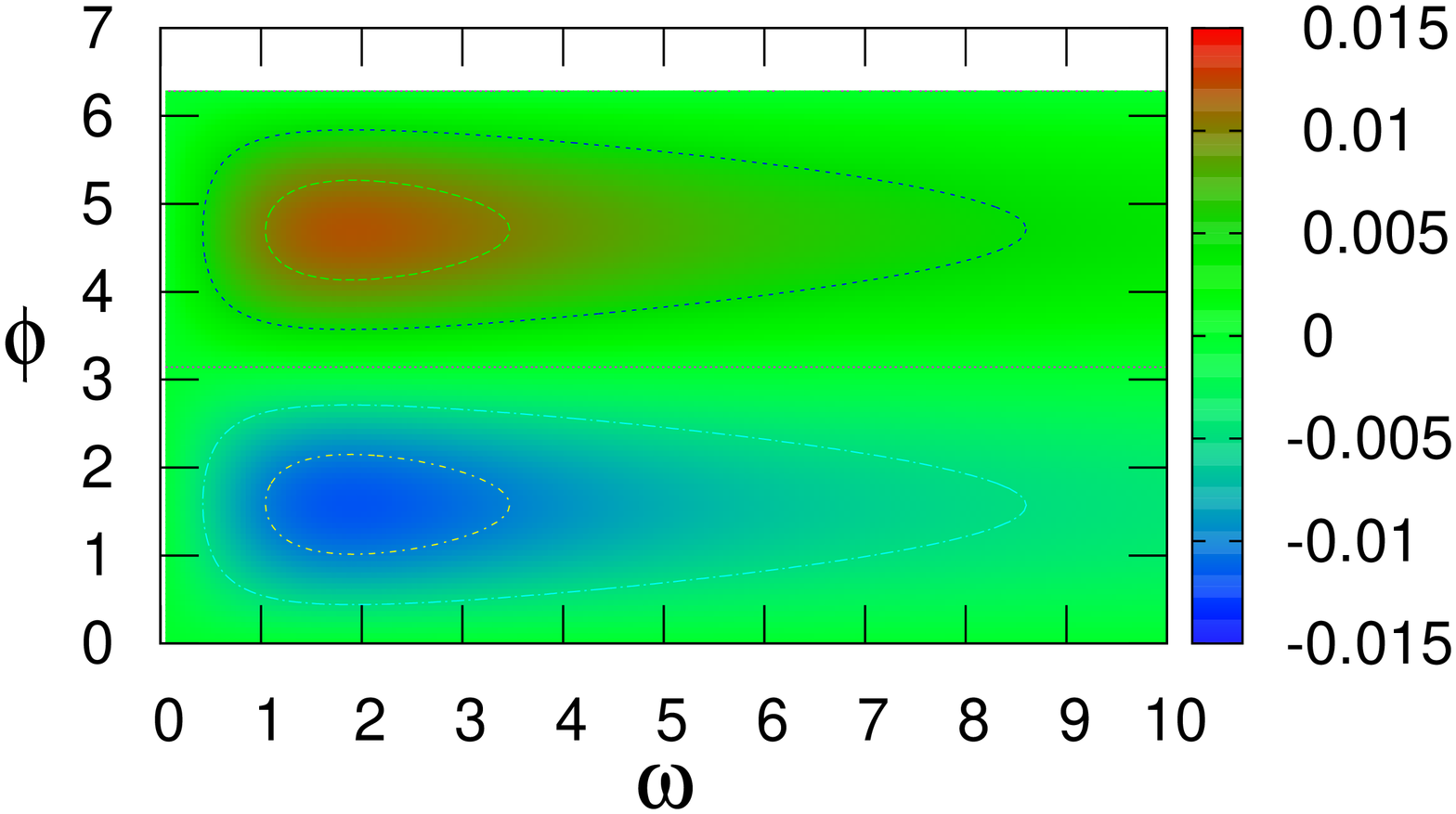}
   \label {c2_2}}
  \subfloat[$N=3$]{
        \includegraphics[angle=-90,width=0.5\linewidth]{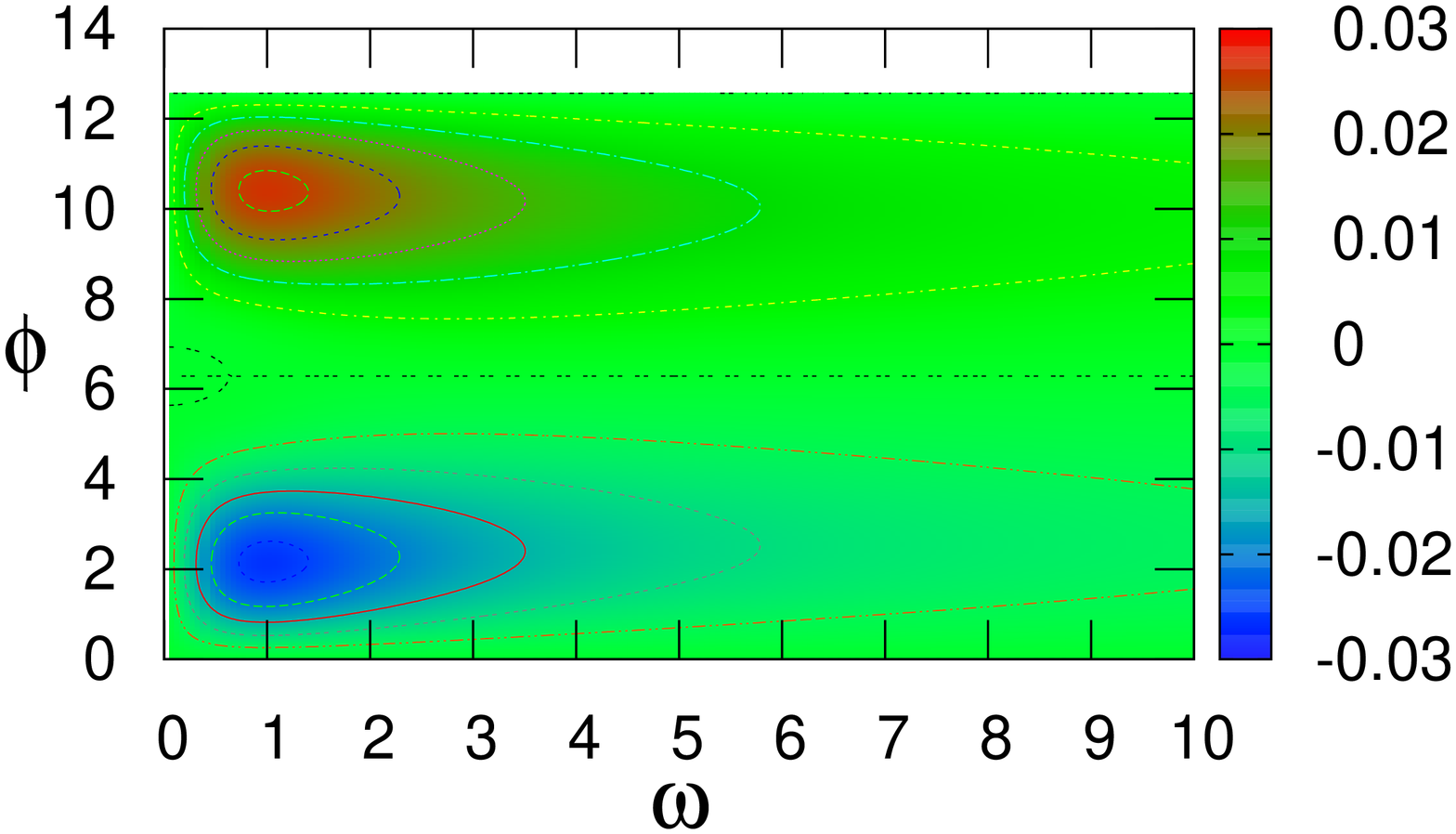}
   \label {c2_3}}

  \subfloat[$N=5$]{
        \includegraphics[angle=-90,width=0.5\linewidth]{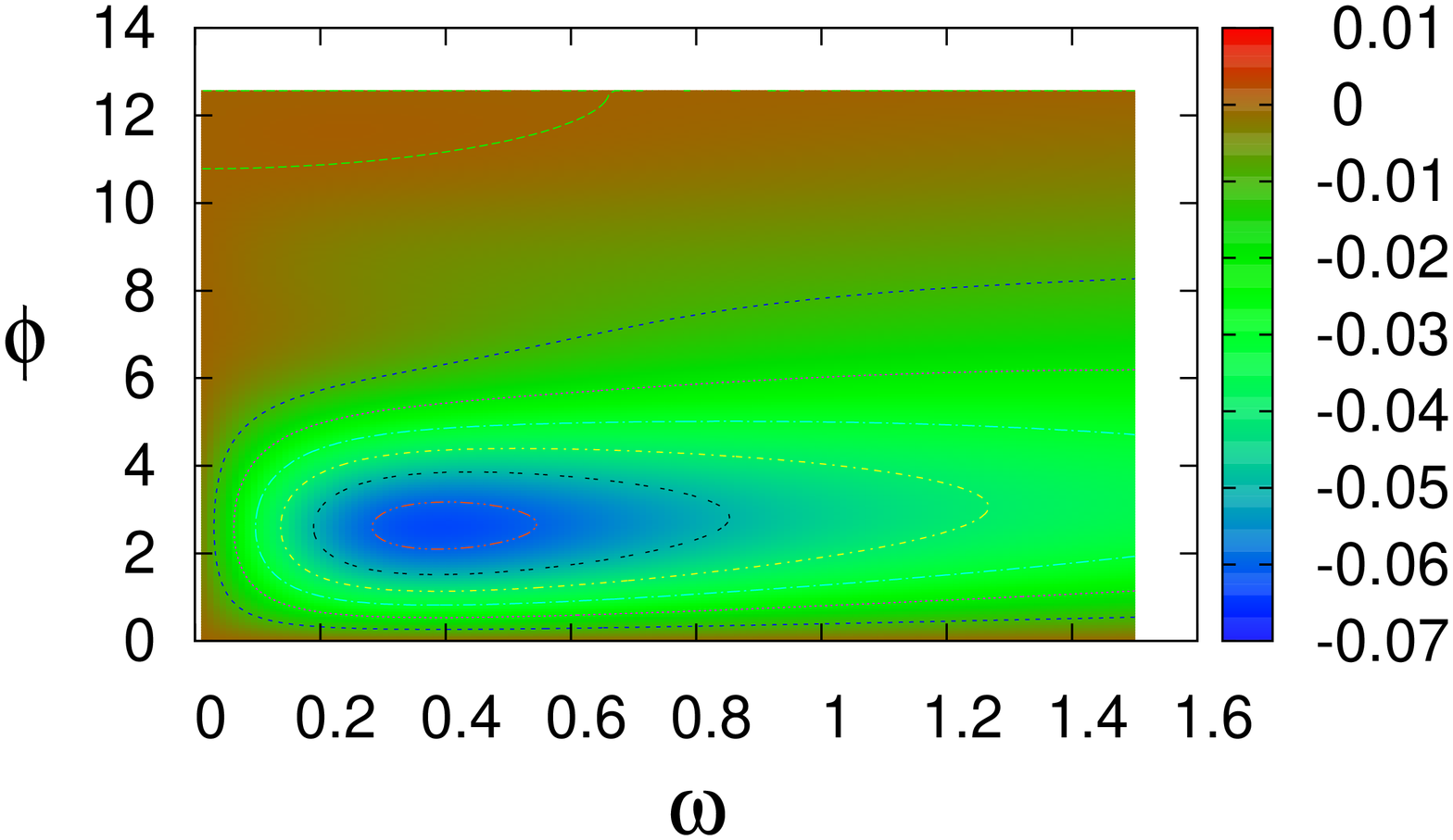}
   \label {c2_5}}
  \subfloat[$N=10$]{
        \includegraphics[angle=-90,width=0.5\linewidth]{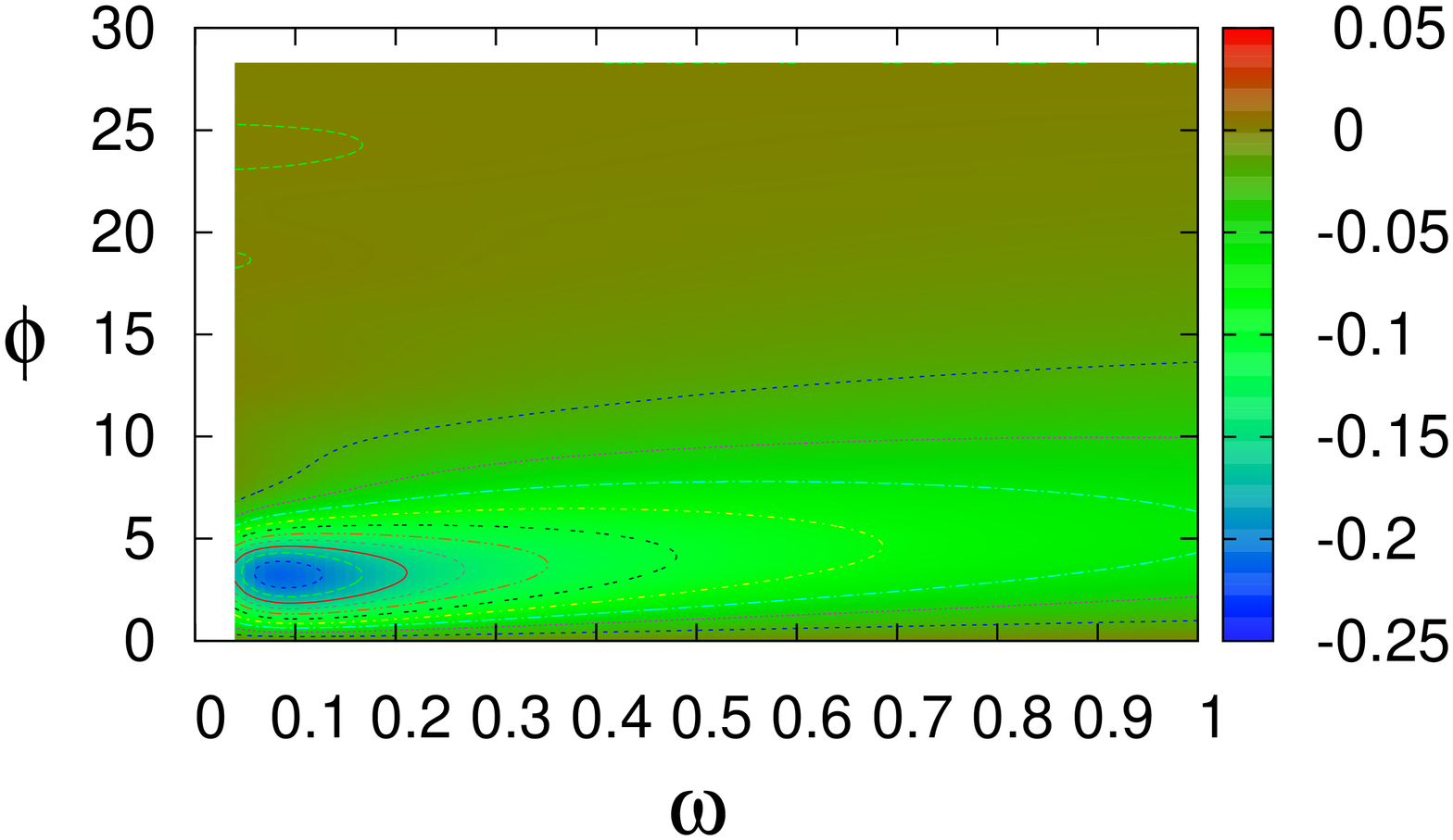}
   \label {c2_10}}

   \caption{(Color online) Frequency ($\omega$) and phase ($\phi$) dependence of the velocity for a different number of particles $N$. Note that for reading purposes, the panels with $N=5$ and 10, have the scale of frequencies different from the others.}
 \label{c2_map}
 \end{figure}

\begin{figure}[tb]
\subfloat[$N=2$]{
        \includegraphics[angle=-90,width=0.5\linewidth]{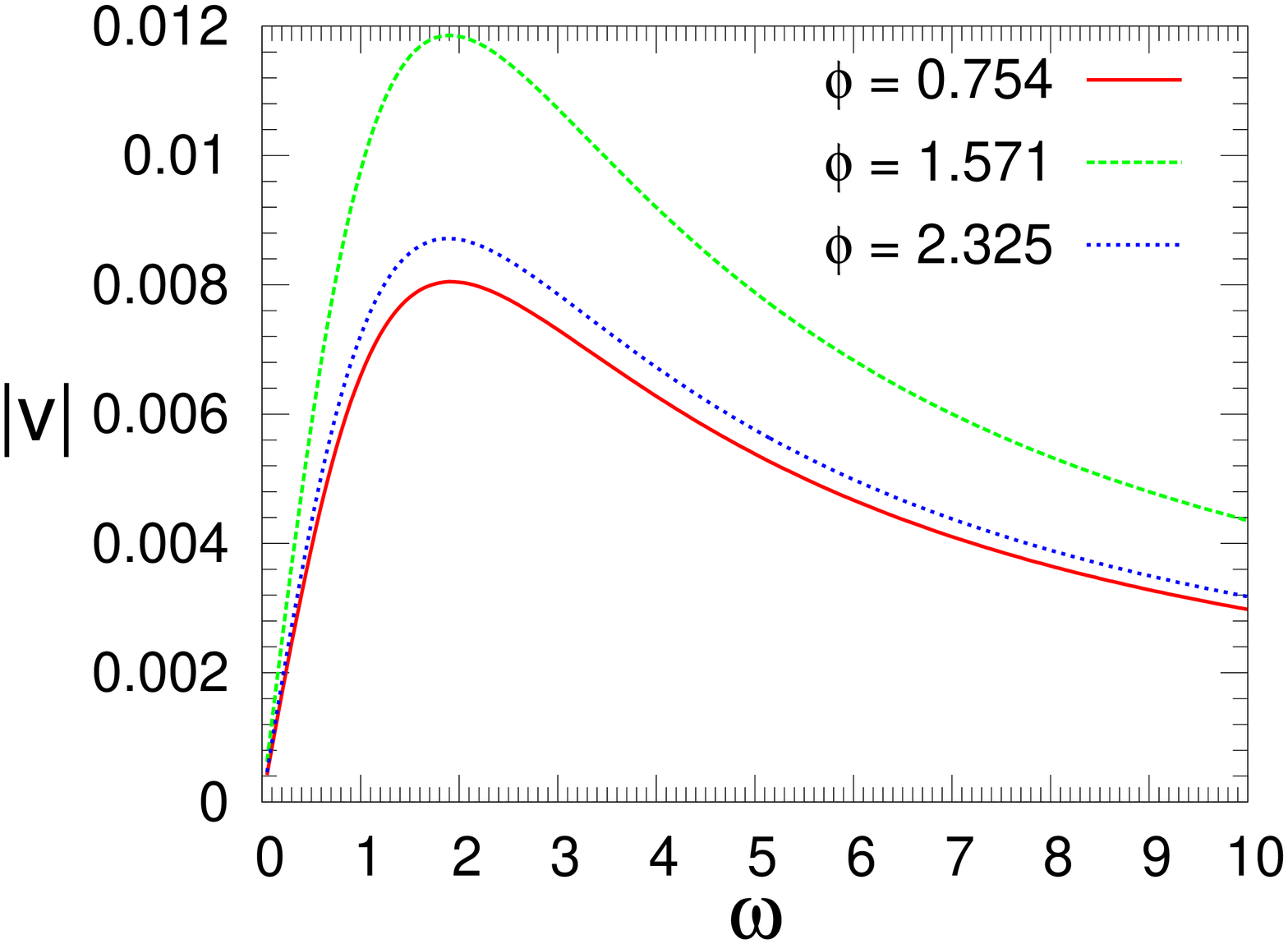}}
\subfloat[$N=3$]{
        \includegraphics[angle=-90,width=0.5\linewidth]{2_frec.eps}}

\subfloat[$N=5$]{
        \includegraphics[angle=-90,width=0.5\linewidth]{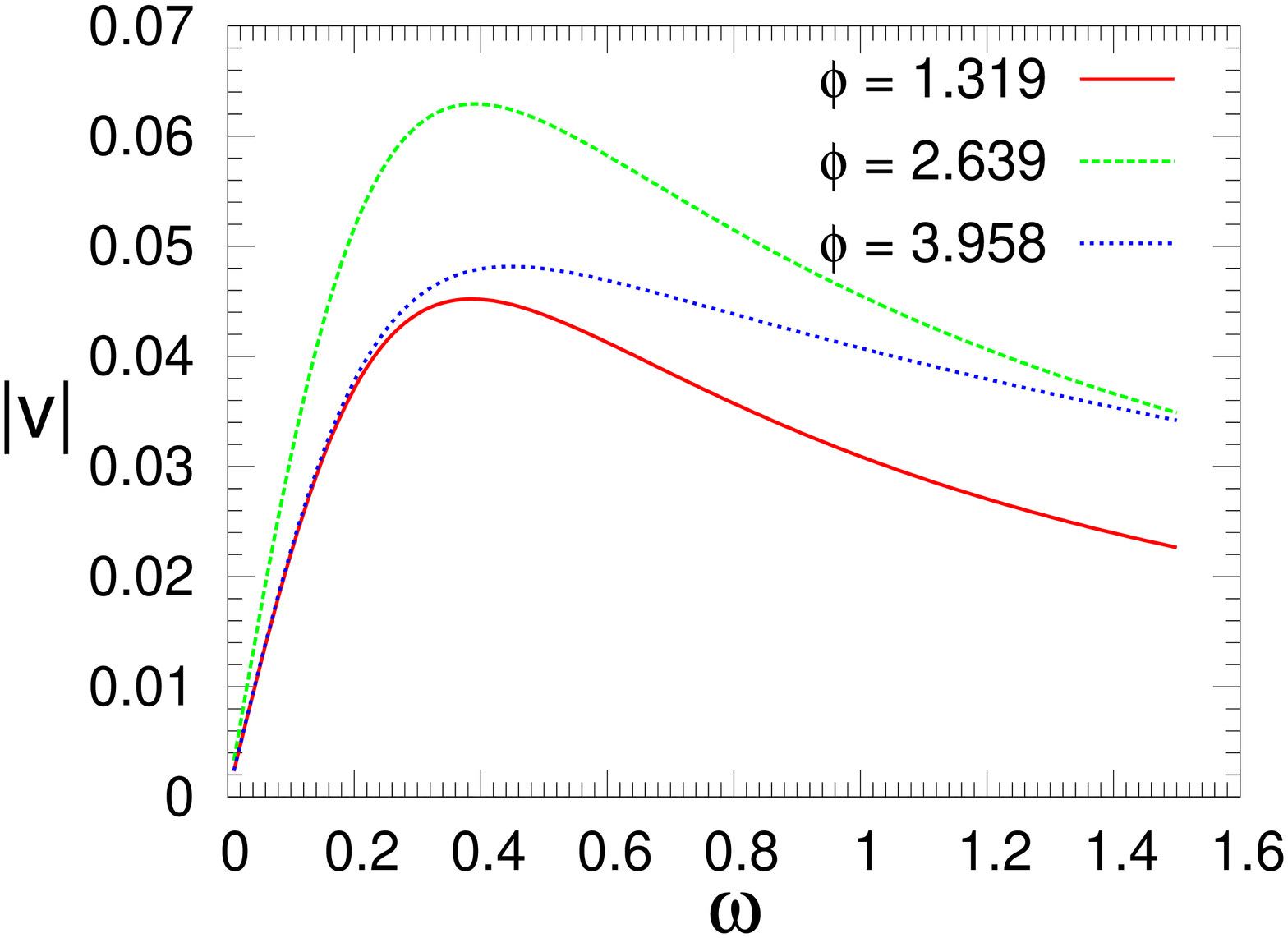}}
\subfloat[$N=10$]{
        \includegraphics[angle=-90,width=0.5\linewidth]{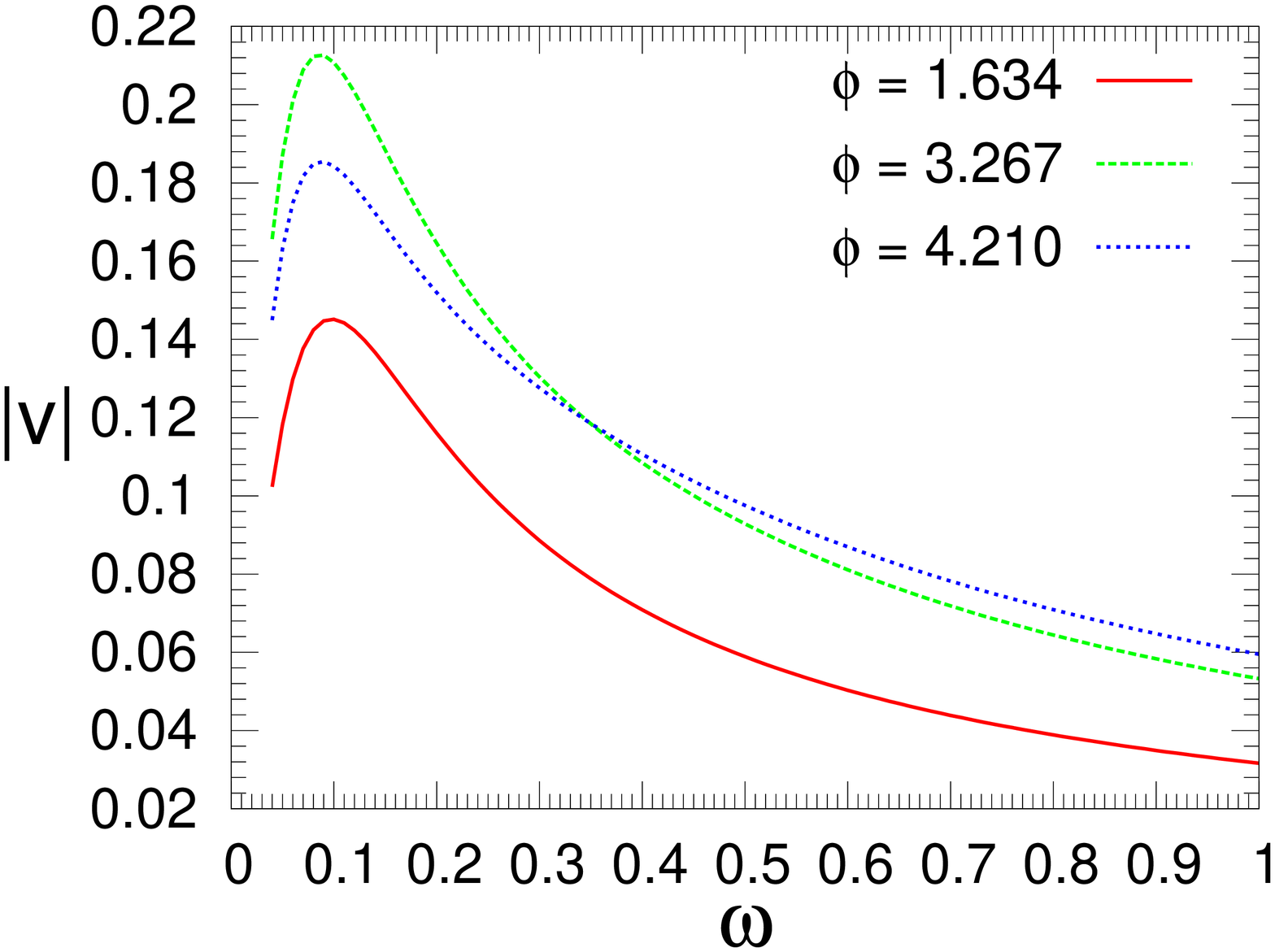}}
\caption{(Color online) Frequency ($\omega$) dependence of the velocity for different number of particles and for three different phases.}
\label{c2_frec}
\end{figure}

We have performed extensive simulations with different numbers of spheres of the polymer. The main measure is the asymptotic velocity (more precisely, the mean velocity over many periods of time) for different values of the driving frequency ($\omega$) and for different values of the total phase difference between the last and the first particle ($\phi$).  Amplitude is fixed to $A = 10$ along all the simulations.
In Fig.~\ref{c2_map} it is summarized the behavior of the asymptotic time average velocity as a function of the two variables say frequency $\omega$ and phase difference  $\phi$, for different values of the chain length, namely $N=2, 3, 5$, and $10$.
 \begin{figure}[t]
\subfloat[$N=2$]
       {\includegraphics[angle=-90,width=0.5\linewidth]{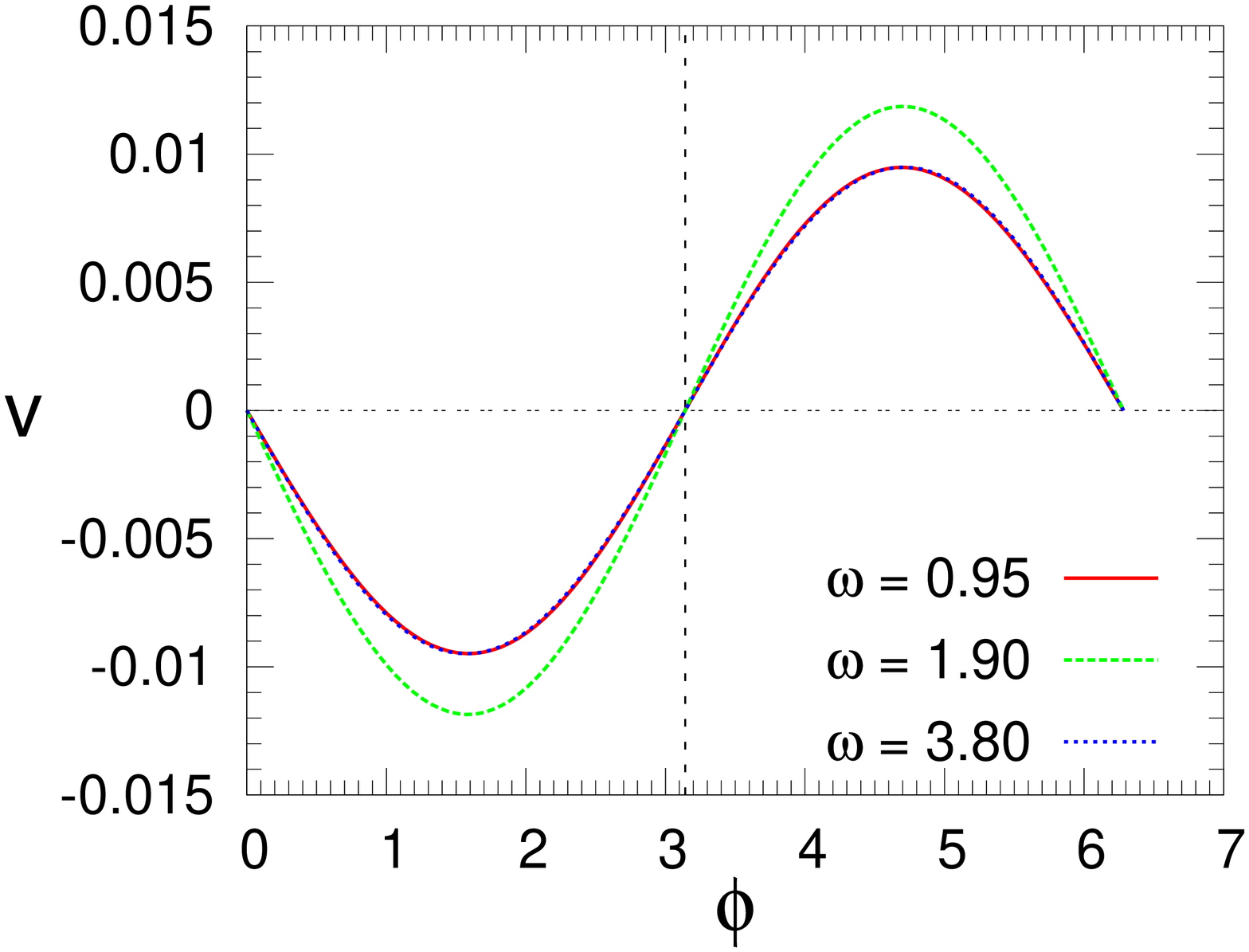}}
\subfloat[$N=3$]
       {\includegraphics[angle=-90,width=0.5\linewidth]{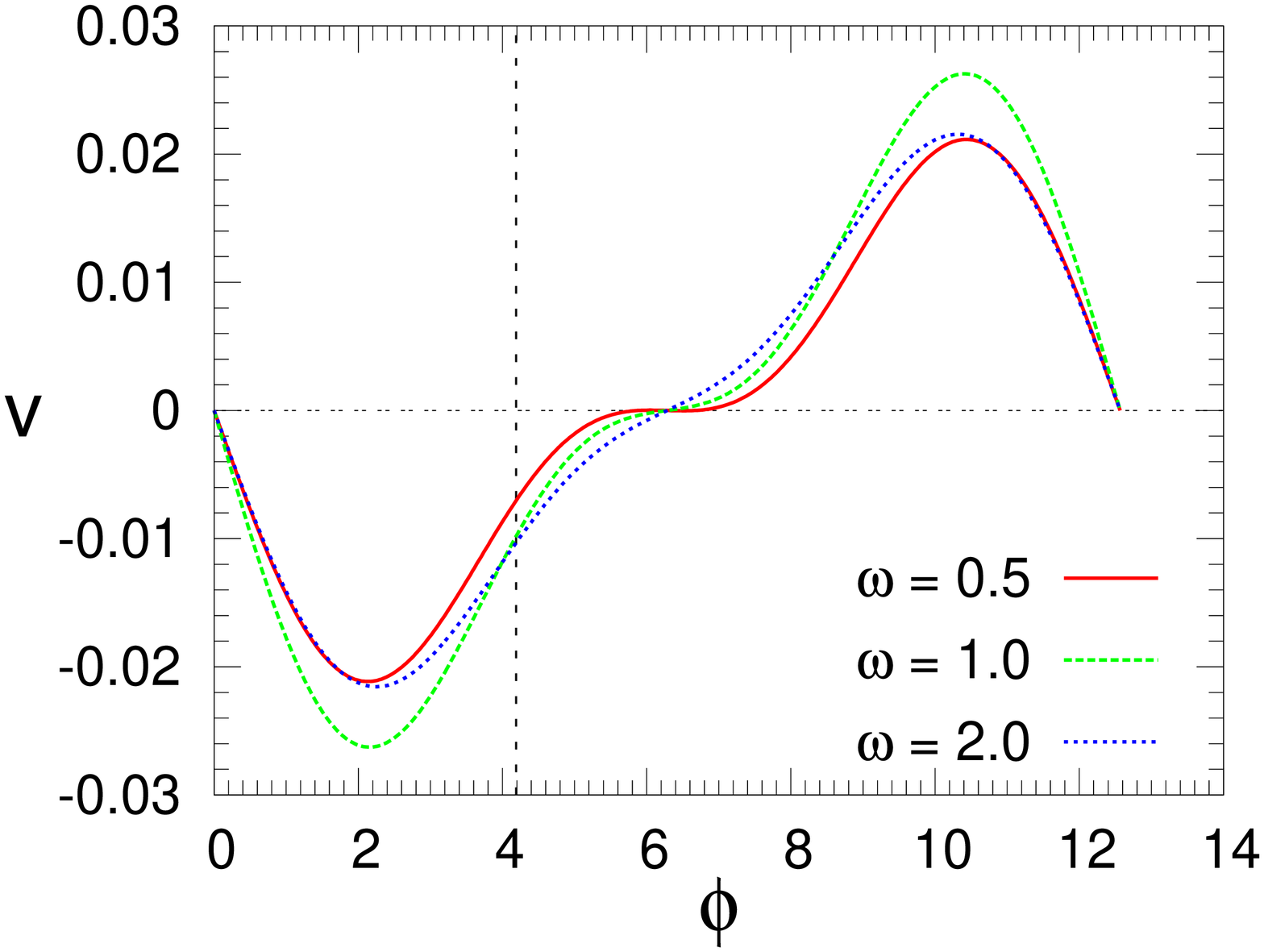}}

\subfloat[$N=5$]{
        \includegraphics[angle=-90,width=0.5\linewidth]{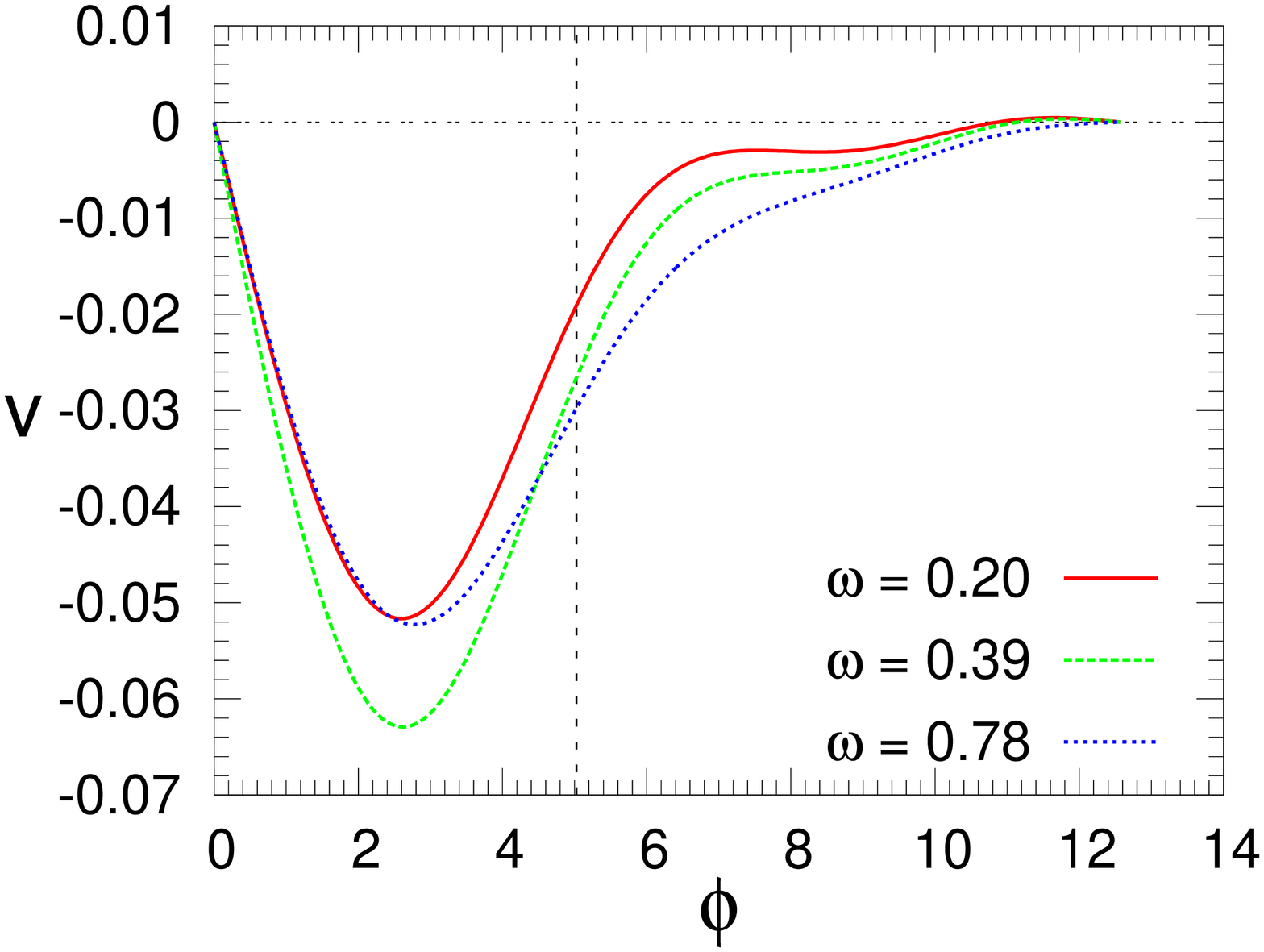}}
\subfloat[$N=10$]{
        \includegraphics[angle=-90,width=0.5\linewidth]{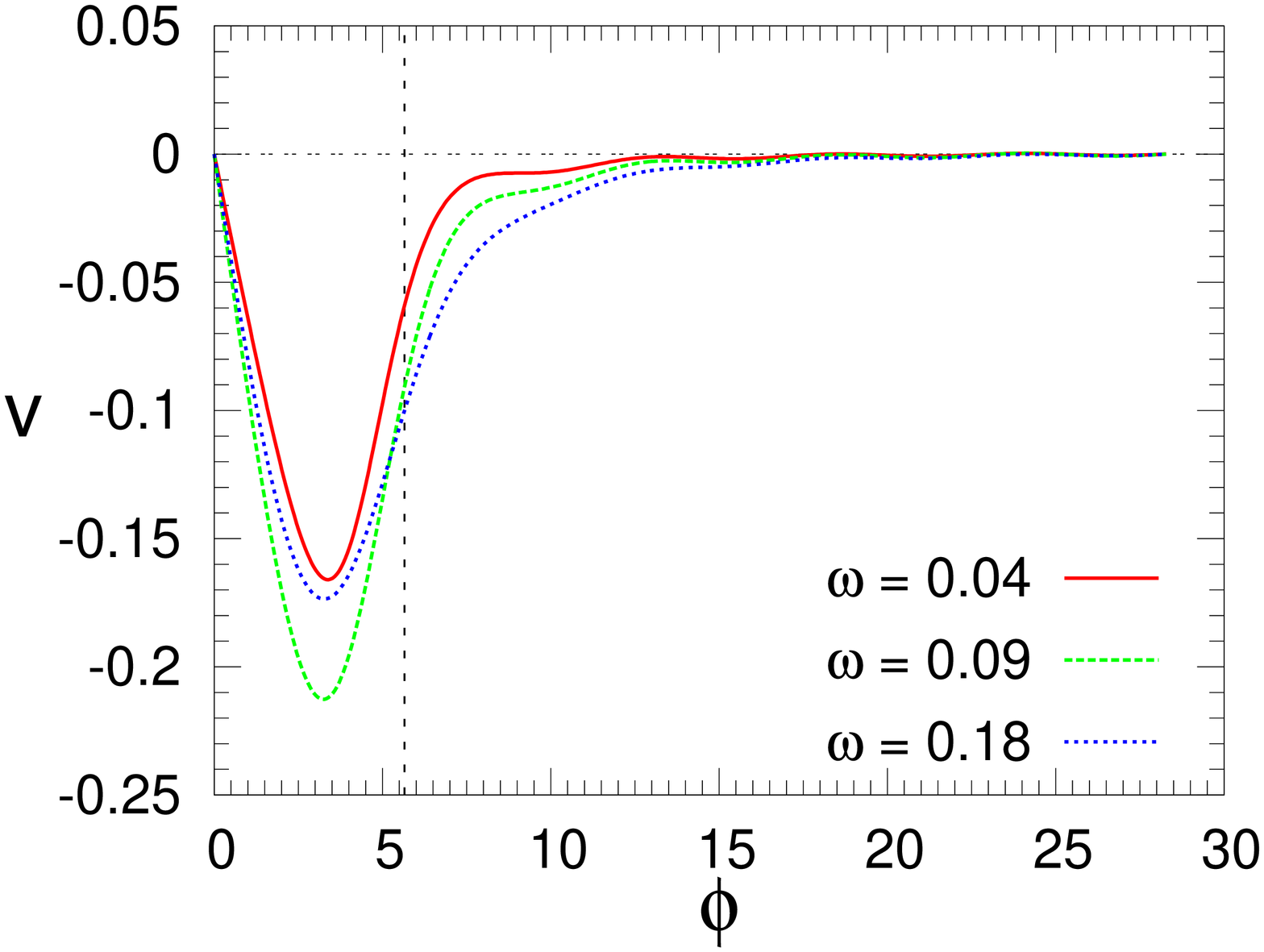}}
\caption{(Color online) Phase difference ($\phi$) dependence of the velocity. The vertical lines represent the first phase difference at which the \emph{external} total force $F_{\rm tot}$ is $0$ at any time. Note that for $N=5$ and 10 only a half period in phase is shown.}
\label{c2_fase}
\end{figure}
We can observe clear maxima and minima of the velocity which strongly depend on the number of particles $N$. In particular, by increasing $N$, the $\omega$-$\phi$ region where the velocity is significantly different from zero reduces rapidly with the frequency, while the absolute values of the velocity maxima increases.
These features are better illustrated in Figs.~\ref{c2_fase} and \ref{c2_frec}, where a few cross sections of the previous plots have been depicted. In Fig.~\ref{c2_fase} we can see that for $N=2$ ((a) panel) a minimum and a maximum appear for the velocities as a function of the phase difference, for fixed values of the frequency. In this case the periodicity in the total phase difference is given by $\phi_P (N=2) = 2\pi$. The case $N=3$ seems more complex, but still with one minimum and one maximum, with a total phase difference $\phi_P(N=3) = 4\pi$. In general the period depends on $N$: $\phi_P = (N-1) 2\pi$ as it has been shown above. For clarity, panels (c) and (d) only show a semi-period. The three frequencies used in all the plots (see the legend in each panel), have been chosen before, after and around the maxima values visible in Fig.~\ref{c2_map}.

The vertical lines present in the Fig.~\ref{c2_fase} indicates the first phase values at which the total external force on the system is null at any time $t$. In that case the movement can be associated with an internal movement only, typically used for swimming, where the external force is zero. It is there visible that a net velocity is found for those values for $N \geqslant 3$, as also found in other works on swimming systems. Applying the sum rule for the sine function, the null external force is recovered for the phase value $\phi = 2\pi (N-1) / N$.

 \begin{figure}[t]
\includegraphics[angle=-90,width=1.0\linewidth]{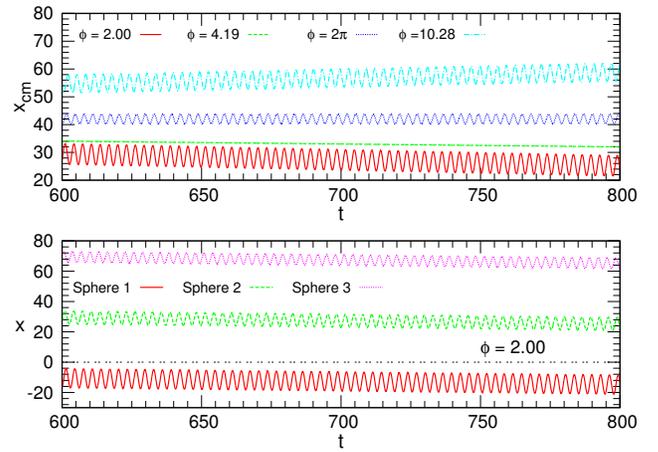}
\caption{(Color online) Upper panel: Trajectories of the center of mass of the polymer for the case $N=3$ and for different values of the total phase $\phi$. Lower panel: Trajectory of each of the three particles for the phase value $\phi = 2.0$. All trajectories are calculated with the frequency $\omega = 2.0$.}
 \label{Traj}
 \end{figure}
To better visualize the kinetic behavior of the chain, a set of trajectories of the center of mass have been drawn in Fig.~\ref{Traj} for the case $N=3$. We can observe in the upper panel that for $\phi=2 \pi$ no displacement is present. The cases $\phi=2.0$ and $\phi=10.28$ represent respectively a value around the minimum and the maximum of the velocity, as visible in the corresponding plot of Fig.~\ref{c2_fase}. The other phase value $\phi=4.29$ lies close to the null external force. In that case we can see that the center of mass does not present visible oscillations (because the oscillating external forces compensate) and the net motion has a negative velocity, as also shown in Fig.~\ref{c2_fase} with $N=3$ at the velocity given at the vertical line. The lower panel of Fig.~\ref{Traj} shows the motion of each of the three spheres for the phase value $\phi = 2.0$. We can observe the broken symmetry in the amplitude of the three different oscillations. The leading particle undergoes the larger oscillations. This is true for all values of the total phase difference independently of the velocity direction.

Figure \ref{c2_frec} shows the cross section of the velocity as a function on the frequency, for different values of the phase difference. The interesting result is that a clear maximum appears for all the chain lengths. This maximum decreases in the frequency value, as we increase the number of spheres in the chain. In the extreme case of a very large number of spheres, we can argue that no movement persists, except at $\omega = 0$ (i.e. for a force constant in time).

It is interesting to note that the null value of the velocity corresponds to the maximum force amplitude to the system. In fact, for example, for $\phi=0$, the force is the same for all the spheres, and the total external force on the chain is simply given by $F_{\rm tot} = N A$, where $A$ is the amplitude of the force on each sphere. The reason why the mean velocity is zero, is that the system is in that case strongly symmetric, and the average over many time averages gives a zero mean velocity.

\begin{figure}[b]
\includegraphics[angle=-90,width=1.0\linewidth]{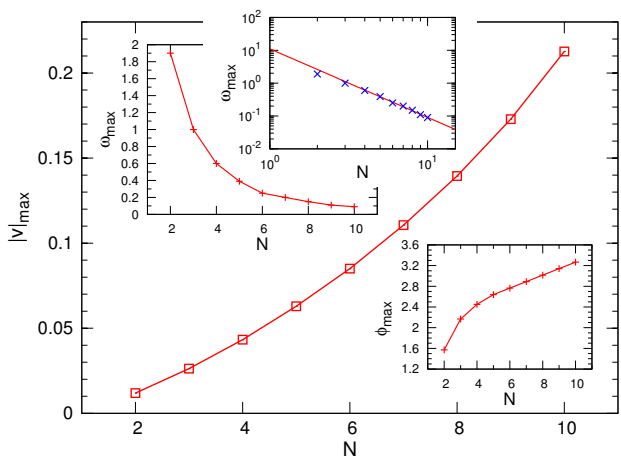}
\caption{(Color online) Modulus of the chain velocity maximum as a function of the number of spheres $N$. Insets: Frequency of the maximum in linear plot (top left) and log-log plot (top right) , and phase of the maximum (bottom).}
 \label{N}
 \end{figure}

Figure \ref{N} shows the behavior of the values of the maxima of the velocity, where we observe their rapid increasing trend as a function of the number $N$ of spheres in the chain. This is due to the fact that the maximum velocity occurs for a net force different from zero, and so an increase of the number of particles results in an increase of the value of the force acting on the chain. The insets of Fig.~\ref{N} show the frequency $\omega_{\rm max}$ and the total phase difference $\phi_{\rm max}$ corresponding to the maximum velocity as a function of the number of spheres $N$. By increasing $N$,  $\omega_{\rm max}$ moves toward the zero value and a power-law scaling $\omega_{\rm max} \approx N^{-1.9}$ is also observed.

Finally, we should stress that this non-monotonic behavior of $v$ with $\omega$ does not appear in the study of self-propulsion system like those of Refs.~\cite{3_esferas_1,3_esferas_2} where a simple $v \propto \omega$ is observed. However, it is recovered in Ref.~\cite{earl2007} when some driving force is applied. This behavior has been observed in the two experiments \cite{dreyfus-nat2005,garstecki2009} cited before on devices (at very different length scales) under external fields.

%
\section{DISCUSSION AND CONCLUDING REMARKS}

This work presents a model of directional motion based on the effect of modulated driving forces and hydrodynamical interaction between particles. The introduction of inversion-symmetry-breaking fields to provoke unidirectional motion has been suggested previously in other contexts, like the motion of molecular motors \cite{cilla2001, niurka}, motion of solitons \cite{flach}, or active Brownian motion \cite{fia1,fia2,fia3}. In these cases it is needed either to have a substrate potential or inertial effects or the presence of fluctuations.
On the contrary, this model deals with a chain in an overdamped environment without any external potential.

It is interesting to stress that the net velocity is given by the hydrodynamical contribution, as it is simple to verify. In fact, if the friction is local, (i.e., the Oseen matrix of Eq.~(\ref{tensor_ossen}) remains with the diagonal terms only) the equation of motion (\ref{eq-motion}) becomes linear, and the sum of the forces (internal and external) over all particles, averages exactly to zero. In this sense, it is clear that the hydrodynamical interactions introduce a non local and non linear contribution which generates the peculiar net velocity here found.
Therefore, this contribution is an essential term to be considered in these kinds of problems.

Another interesting point to discuss is the fact that, in our system, the two particle chain presents net motion for a large range of parameters. This could appear in contradiction with the so called 'scallop theorem" \cite{purcell, LaugaSoft2011}, which requires a system with, at least, two degrees of freedom to support the motion. However, one has to note that our system does not fulfill the other requirement of the 'scallop theorem". In fact, in that case, deformation of the swimming object should be driven by internal forces that average to zero at \emph{any time}. This only occurs here for certain total phase differences $\phi$, whose values are marked with a vertical line in Fig.~\ref{c2_fase}. In fact, for $N=2$, the theorem hypothesis (null force) is fulfilled for $\phi = \pi$, and the velocity is zero at this point. Larger polymers ($N>2$) show net velocity at these special null-force points.

In summary, we analyze a polymer chain which moves in a low Reynolds number regime under a driving force which depends on the positions of the particles it is composed of. The study is mainly devoted to the influence of the driving force to the possible presence of the net velocity in the system. The presence of some symmetries in the equations has been analyzed as well as the relationship of the chain velocity with the total phase difference between the last and the first particles of the chain, finding a maximum, and revealing that the directionality can be changed by controlling the phase differences of the driving force across the chain. The maximum of the velocity is also present as a function of the frequency of the applied force, showing again an optimal value for the transport which depends on the number of particles in the chain. The model here presented includes a special case of self propelled swimming, given specifically by a total phase difference such that the total external force acting on the chain be null. In that case the results are in agreement with other studies. The velocity behavior predicted by this model is also in agreement with the experimental outcomes of time dependent driving forces on extended bodies.

This study is a first approach to the effect of both a time and a spatial dependent field acting on elastically deformable objects. The driving proposed here can be performed in devices at microscopic scale using electromagnetic waves where phase differences are related to the particle position. This means that it could be possible to use a suitable light beam to drive the polymer. Our model is a simple approximation to this experimental proposal if the particle oscillations are small in comparison to the distance among them. The possibility of tuning the value and direction of the velocity with few parameters makes this system feasible to pilot structures in viscous environments.

\begin{acknowledgments}
This work is supported by the Spanish DGICYT Projects No. FIS2008-01240, and No. FIS2011-25167, both co-financed by FEDER funds. Financial support from European Science Foundation Research Network "Exploring the Physics of Small Devices", and the critical reading by L. M. Flor\'ia and J. J. Mazo are also acknowledged.
\end{acknowledgments}

\end{document}